%% file: main.tex
\documentclass[conference]{IEEEtran}
\IEEEoverridecommandlockouts

\usepackage{cite}
\usepackage{amsmath,amssymb,amsfonts}
\usepackage{algorithmic}
\usepackage{graphicx}
\usepackage{textcomp}
\usepackage{xcolor}
\usepackage{algorithm}
\usepackage{color, soul}
\usepackage{comment}
\usepackage{multicol}
\usepackage{multirow}
\usepackage{adjustbox}
\usepackage{tablefootnote}
\usepackage{threeparttable}

\def\BibTeX{{\rm B\kern-.05em{\sc i\kern-.025em b}\kern-.08em
    T\kern-.1667em\lower.7ex\hbox{E}\kern-.125emX}}
\begin{document}

\title{Masked Autoencoders as Universal Speech Enhancer

\thanks{*Work performed during internship at AWS AI Labs.}
}

\author{\IEEEauthorblockN{Rajalaxmi Rajagopalan*}
\IEEEauthorblockA{\textit{University of Illinois, Urbana-Champaign}}
\and
\IEEEauthorblockN{Ritwik Giri}
\IEEEauthorblockA{\textit{AWS AI Labs}}
\and
\IEEEauthorblockN{Zhiqiang Tang}
\IEEEauthorblockA{\textit{AWS AI Labs}}
\and

\IEEEauthorblockN{Kyu Han}
\IEEEauthorblockA{\textit{AWS AI Labs}}
\and
\IEEEauthorblockN{Gerald Friedland}
\IEEEauthorblockA{\centerline{\textit{AWS AI Labs}}}
}

\maketitle

\input{tex/abs_intro}

\input{tex/rest}

\bibliographystyle{IEEEtran}
\bibliography{IEEEabrv,strings,refs}

\end{document}

%% file: tex/abs_intro.tex
\begin{abstract}
Supervised speech enhancement methods have been very successful. However, in practical scenarios, there is a lack of clean speech, and self-supervised learning-based (SSL) speech enhancement methods that offer comparable enhancement performance and can be applied to other speech-related downstream applications are desired. In this work, we develop a masked autoencoder based universal speech enhancer that is agnostic to the type of distortion affecting speech, can handle multiple distortions simultaneously, and is trained in a self-supervised manner.
An augmentation stack adds further distortions to the noisy input data. The masked autoencoder model learns to remove the added distortions along with reconstructing the masked regions of the spectrogram during pre-training. The pre-trained embeddings are then used by fine-tuning models trained on a small amount of paired data for specific downstream tasks. We evaluate the pre-trained features for denoising and dereverberation downstream tasks. We explore different augmentations (like single or multi-speaker) in the pre-training augmentation stack and the effect of different noisy input feature representations (like $log1p$ compression) on pre-trained embeddings and downstream fine-tuning enhancement performance. We show that the proposed method not only outperforms the baseline but also achieves state-of-the-art performance for both in-domain and out-of-domain evaluation datasets. 
\end{abstract}
\section{Introduction}
\label{sec:intro}
Speech enhancement involves the reconstruction of clean target speech from a noisy recording corrupted by ambient noise, interfering speech, distortions, etc. 
Deep learning approaches \cite{hao2021fullsubnet,defossez2020real,zhao2018two,zhao2018convolutional,koizumi2020speech,isik2020poconet,pandey2021dense,richter2023speech,fu2019metricgan,pascual2017segan,choi2018phase} have successfully outperformed classical ones especially for denoising.
More recently, there has been a paradigm shift from isolated models for each speech task to pre-trained models that learn universal useful speech representations that improve various downstream tasks. The pre-training-finetuning model framework is especially attractive when the pre-trained model is self-supervised like HuBERT \cite{hsu2021hubert}  and WAVLM \cite{chen2022wavlm}. Audio classification tasks like speaker identification, speech recognition, etc., and audio enhancement tasks like denoising, dereverberation, etc., are the two major classes of downstream tasks. Enhancement downstream tasks have derived less benefit from general pre-trained models. For instance, pre-trained models applied to classification tasks like keyword spotting, speaker identification, etc  \cite{niizumi2022masked,chong2023masked} have shown great performance but not for audio enhancement. 
There is increasing interest in improving pre-trained models for downstream enhancement \cite{song2023exploring,zhong2023extending,kong2021speech}. 
Thus, in this work we develop a Masked Autoeconder (MAE) based model pre-trained in a self-supervised fashion that (1) can tackle a broad range of distortions simultaneously, (2) trained without access to large-scale clean speech corpus, and (3) offers robust performance for enhancement downstream tasks.

%% file: tex/rest.tex
\section{RELATED WORK}
\label{sec:related}
 The majority of speech enhancement approaches rely on fully supervised learning that requires high-quality clean speech reference which is difficult to obtain. 
 Thus, self-supervised approaches that will relax this constraint 
 are attractive. There have been overwhelmingly successful SSL approaches such as HuBERT \cite{hsu2021hubert} and WavLM \cite{chen2022wavlm} that develop pre-trained models for various downstream audio tasks like speaker identification, separation, and diarization. An open area of active research is exploring the viability of such SSL pre-trained models for universal speech enhancement. 

WavLM and HuBERT employ a classification/prediction-based objective that is mismatched with the continuous nature of the enhancement task; estimating continuous enhanced/restored speech as highlighted by works like \cite{song2023exploring}. Results show limited success in extending pre-trained models for enhancement and cannot beat the performance of SOTA discriminative models like Phase-aware U-Net SE \cite{choi2018phase} and SOTA diffusion models \cite{richter2023speech,serra2022universal}.
Since our primary goal is to develop a foundation model for speech enhancement, we instead 
use regression loss similar to Masked Autoencoder \cite{huang2022masked} where some portions of the input noisy STFT are masked and a vision transformer-based (ViT) model learns to reconstruct the masked regions thereby learning robust speech representations. Authors of \cite{zhong2023extending} illustrate the benefit of Masked Autoencoder pre-training on denoising fine-tuning. However, they do not consider other distortions like background interfering speech, clipping, codec artifacts, etc, during pretraining, and hence neither their pre-trained model nor the downstream fine-tuned model is equipped to handle other distortions. 

In contrast, we propose a universal self-supervised speech enhancement framework that builds on the masked autoencoder \cite{huang2022masked} by adding an augmentation stack that adds more distortions like distance-based multi-speaker speech mixtures, reverberation, clipping, codec artifacts, etc, to the general noisy speech input and the pre-trained model learns to both reconstruct masked portions and remove the added distortions. This dual objective ensures that the pre-trained model is \textbf{self-supervised} and \textbf{universal}. For instance unlike previous works \cite{zhang2021restoring}, multi-speaker augmented pre-training will learn features that enable removal of other undesired \textit{speech like} noise such as background babble, a TV in the background playing intelligible speech, or background interfering speakers farther from the target speaker. Thus, this pre-training scheme offers robust embeddings that can be further finetuned with a small amount of labeled data for different downstream tasks like denoising, dereverberation, source separation, bandwidth extension, etc.

\begin{figure*}
\centering
{\includegraphics[width=1.8\columnwidth]{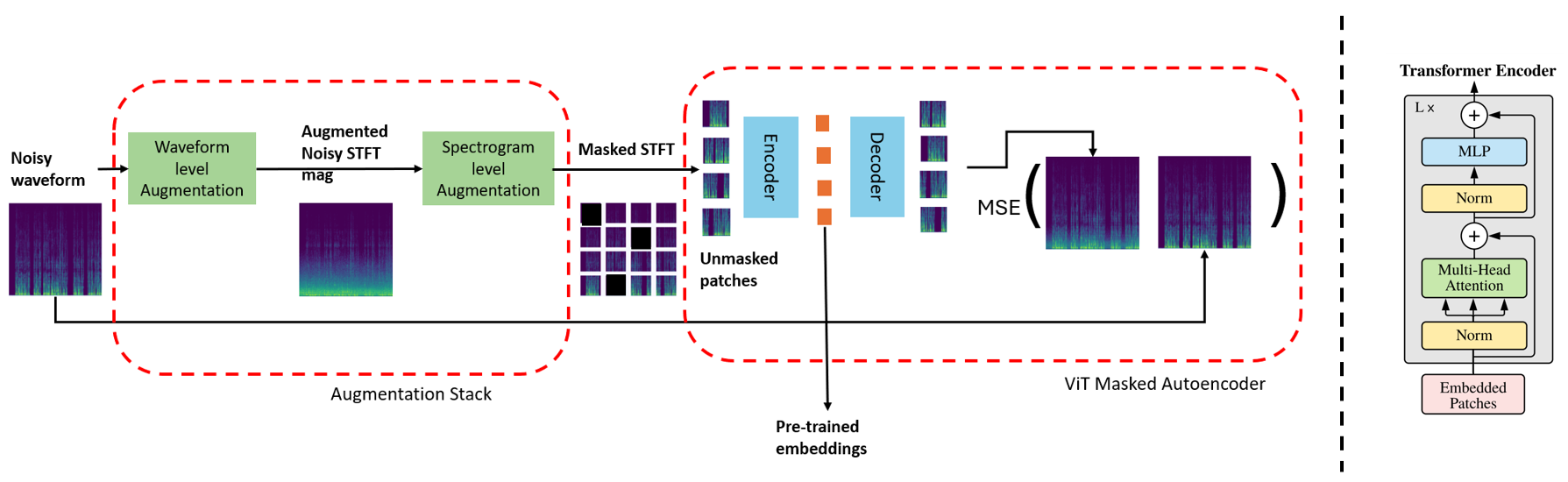}}
\vspace{-0.2in}
\caption{\textit{Pretraining framework}: (a) Augmentation stack (b) ViT Masked Autoencoder \cite{alexey2020image}}
\label{fig:pretrain}
\vspace{-0.1in}
\end{figure*}  

\section{PROPOSED METHOD}
\label{sec:method}
For a speech clip $s \in \mathbf{R}^L$, the speech distortion process is modeled as a function $d(\cdot)$ and the degraded speech $x \in \mathbf{R}^L$, $x = d(s)$. Speech enhancement is an inverse problem that aims to restore high-quality speech $\hat{s} = f(x)$ from $x$, where $f(\cdot)$ is the enhancement function; an inverse approximation of $d(\cdot)$. 
The various distortions impacting $x$ is a composite function,
\vspace{-0.1in}
\begin{equation}
\begin{aligned}
d(x) &= d_1 \circ d_2 \circ \dots d_Q(x) \in \mathbf{D} 
\end{aligned}
\label{eqn:distort}
\vspace{-0.05in}
\end{equation}
where $\circ$ denotes function composition, $\mathbf{D} = {d_v(\cdot)}_{v=1}^V$ is the set of distortion types, $Q \leq V$ is the number of distortions in $d(\cdot)$. 
In our SSL universal enhancement framework, 
we first pre-train a general model in a self-supervised manner on large amounts of noisy speech to learn speech representations that are then used by specific fine-tuning models trained on small amounts of paired clean-noisy speech for downstream tasks. 

\subsection{Pretraining}
\label{sec:pretrain}
Figure \ref{fig:pretrain} shows two main modules of the pre-training:
\begin{itemize}
    \item[(1)] \textbf{\textit{Augmentation Stack}}: The general noisy speech is further distorted by adding several distortions in the time domain waveform level (background noise, reverberation, codec artifacts, etc) and STFT spectrogram level (masking).
    \item[(2)] \textbf{\textit{Masked Autoencoder (MAE)}}: A vanilla ViT-AE \cite{huang2022masked} model with encoder and decoder takes the augmented noisy STFT magnitude as input and reconstructs the speech before augmentation was added.
\end{itemize}
This is a self-supervised learning approach as no clean reference is required and by injecting distortion, the noisy waveform before augmentation serves as the reference.

\setlength{\textfloatsep}{0pt}
\begin{algorithm}
\caption{Distance-based Mixture Generation: 2 speakers}\label{alg:multi}
\begin{algorithmic}[1]
\STATE{Target speaker: $s_1$, Interfering speaker: $s_2$}
\STATE{To ensure all speakers are in the same room: pick RIR $r_0$.}
\STATE{First, $s_1$ must not too far away from the microphone}
\STATE{Second, $s_1$ must be closer to the microphone than $s_2$}
\STATE{Compute the Direct-Reverberant ratio (DRR) of $r_0$; 
\vspace{-0.1in}
\begin{equation}
\begin{aligned}
\text{DRR(dB)} = 10*log\left(\frac{P_D}{P_R}\right)
\end{aligned}
\label{eqn:drr}
\vspace{-0.05in}
\end{equation}

where, $P_D, P_R$ are the direct path and reverberant powers.}

\IF{DRR $\ge$ Threshold}
\STATE{Apply $r_0$ to $s_1$, $s_1$ is at appropriate distance from mic}
\STATE{Attenuate the direct path $+$ early reflections of $r_0$: $r_0'$}
\STATE{Apply $r_0'$ to $s_2$, $s_2$ is farther away from mic than $s_1$}
\ELSE
\STATE{Apply $r_0$ to $s_2$, $s_2$ is far from the mic}
\STATE{Decay $r_0$'s late reverberant part using a decay function 
\vspace{-0.1in}
\begin{equation}
\begin{aligned}
A(t) = 
     \begin{cases}
       1 &\quad t < T_0\\
       \frac{1+\alpha}{2} + \frac{1-\alpha}{2} \text{cos} \frac{\pi(t-T_0)}{(T_1-T_0)} &\quad T_0 < t < T_1 \\
       \alpha &\quad t> T_1\\
     \end{cases}
\end{aligned}
\label{eqn:decay}
\vspace{-0.05in}
\end{equation}
$T_0$ and $T_1$ are the RIR's reverberant part that is decayed.}
\STATE{Apply new $r_0''$ to $s_1$, $s_1$ is now closer to mic than $s_2$}
\ENDIF
\end{algorithmic}
\end{algorithm}

\subsubsection{Augmentation Stack}\hfill\\
\noindent\textbf{\textit{Waveform level augmentations:}}

\noindent\textbf{Additive Noise}: Adding noise $n \in \mathbf{R}^L$ to speech $s$:
\vspace{-0.1in}
\begin{equation}
\begin{aligned}
d(s) &= s + n
\end{aligned}
\label{eqn:distort_add}
\vspace{-0.1in}
\end{equation}

\noindent\textbf{Reverberation}: Modeled by convolving speech signals with a Room Impulse Response (RIR) filter $r$:
\vspace{-0.1in}
\begin{equation}
\begin{aligned}
d(s) &= s * r
\end{aligned}
\label{eqn:distort_reverb}
\vspace{-0.1in}
\end{equation}

\noindent\textbf{Clipping}: When the single amplitude exceeds a range [-$\gamma$, +$\gamma$], it is clipped: 
\vspace{-0.13in}
\begin{equation}
\begin{aligned}
d(s) &= \text{max}(\text{min}(s, \gamma), - \gamma), \quad \gamma \in [0, 1]
\end{aligned}
\vspace{-0.1in}
\label{eqn:distort_clip}
\end{equation}

\noindent\textbf{Multi-Speaker Mixtures}: Interfering speech from different speakers is added to the target speaker $s$. Algorithm \ref{alg:multi} shows how modifying RIRs applied to each speaker results in the generation of distance-separated speaker mixtures. Unlike WavLM which generates unrealistic multi-speaker mixtures by ensuring the target utterance is longer compared to inferring speaker (not often observed in practice), distance-separated speaker mixtures are very common in practical scenarios where target speech is corrupted by background babble, background interfering speakers who are not part of the conversation, a TV playing speech in the background, etc. 
\vspace{-0.1in}
\begin{equation}
\begin{aligned}
d(s) &= \mathbf{A} \begin{bmatrix} s * r_0 & s_1 * r_1 & \dots & s_M *r_M \end{bmatrix}^T
\end{aligned}
\vspace{-0.1in}
\label{eqn:multi-spk}
\end{equation}
where, $\mathbf{A}$ is the mixing matrix, $r_i, i \in \{0,1,\dots,M\}$ are the RIRs for each speaker. 
This distance-based cue allows the model to get around the permutation issue between speakers and does not require Permutation-Invariant Training (PIT). Hence the pre-trained model without any finetuning can also be viewed as an enrollment-free alternative approach to target voice extraction, where the target voice is assumed to be closest to the microphone with the highest DRR.

\vspace{0.05in}
\noindent\textbf{\textit{Spectrogram level augmentations:}}
\vspace{0.05in}

\noindent\textbf{Time masking}: $x\%$ of time frames are randomly masked to mimic packet loss concealment \cite{huang2022masked} distortion whereby some portions of the audio clip are lost.

\noindent\textbf{Frequency masking}: $x\%$ of high-frequency bins of the spectrogram are masked. This mimics bandwidth limitation distortion in audio recordings caused by a low sampling rate or defects in the recording device. 

\noindent\textbf{Random TF Masking}: $x\%$ of spectrogram TF bins are randomly masked enabling self-supervision. When a spectrogram portion is masked with a probability $p$, it reduces input sequence length and encourages learning global, contextualized representations from limited unmasked patches.

\begin{figure}[!htp]
\centering
{\includegraphics[width=0.32\columnwidth]{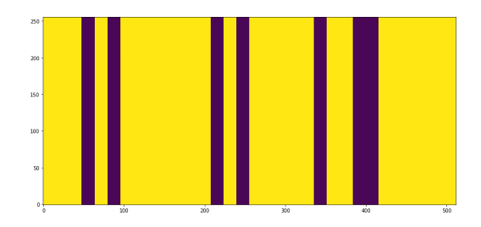}}
\hfill
{\includegraphics[width=0.32\columnwidth]{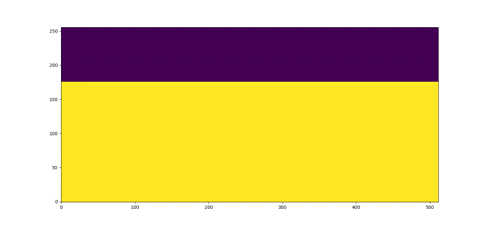}}
\hfill
{\includegraphics[width=0.32\columnwidth]{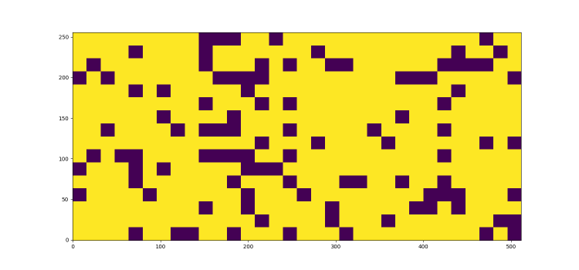}}
\vspace{-0.05in}
\caption{\textit{Spectrogram augmentations}: (a) Time (b) Frequency (c) Random TF}
\label{fig:mask}
\end{figure}

Figure \ref{fig:mask} illustrates the masks that are applied as a dot product to the input STFT magnitude. 
\vspace{-0.15in}

\vspace{-0.1in}
\subsubsection{Masked Autoencoder}
Audio-MAE uses a stack of standard Transformers 
for its encoder and decoder. The encoded patches are padded with trainable masked tokens and passed to the decoder. 
The decoder processes the embeddings and mask tokens to reconstruct the input. The decoder enables the encoder to be trained only on unmasked patches, thereby reducing the computational cost. Global self-attention in image-based MAEs is not applicable to audio spectrograms as the time-frequency context is largely local. 
Thus, Audio-MAEs use local self-attention mechanisms through shifted windowing using SwinTransformer decoder blocks \cite{liu2021swin}.

\vspace{-0.15in}
\begin{figure}[!htp]
\centering
{\includegraphics[width=1\columnwidth]{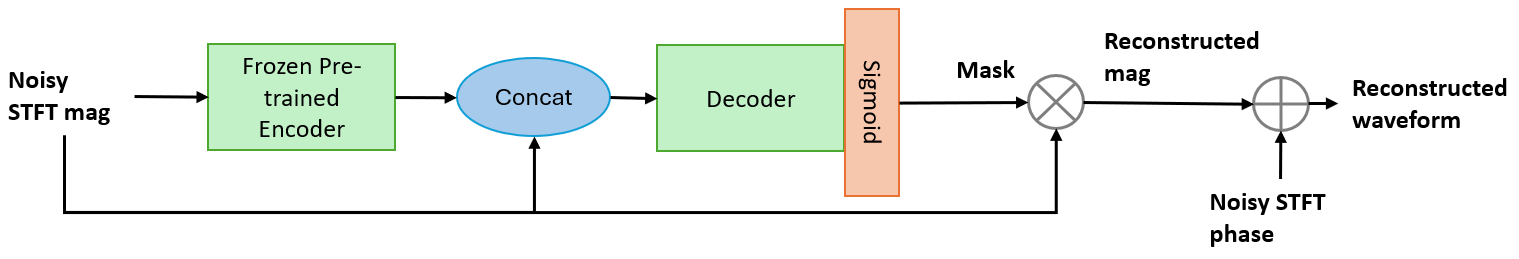}}
\vspace{-0.25in}
\caption{\textit{Finetuning}: Enhancement}
\label{fig:fine}
\vspace{-0.2in}
\end{figure}

\begin{table*}
\centering
\begin{adjustbox}{width=1.6\columnwidth}
\begin{tabular}{c c c c c c c c c }
\hline
Pre-train augmentation type & Model & Finetune & CSIG & CBAK & COVL & PESQ & SSNR & NISQA\\ \hline
- & Noisy & - & 3.34 & 2.44 & 2.63 & 1.97 & 1.68 & 3.05\\ \hline
- & DCUNet \cite{choi2018phase}  & - & 4.24 & \textbf{4.00} & 3.69 & 3.13 & - & -\\ \hline
- & UNIVERSE \cite{serra2022universal} & - & 4.03 & 3.11 & 3.46 & 2.9 & - & -\\ \hline
- & SGMSE+ \cite{richter2023speech}  & - & 4.25 & 3.4 & 3.61 & 2.94 & - & \textbf{4.56}\\ \hline
- & VoiceFixer \cite{liu2021voicefixer} & - & - & - & - & 2.43 & - & -\\ \hline
- & PANN w/ UNet \cite{kong2021speech} & - & 2.43 & 2.96 & 2.3 & 2.28 & - & -\\ \hline
- & Huang et al \cite{huang2022investigating}  & - & - & - & - & 2.68 & - & -\\ \hline
- & Boosting SSL \cite{hung2022boosting} & Y & 4.4 & 3.52 & 3.74 & 3.05 & - & - \\ \hline
Random TF mask & \textbf{MAE for restoration}$^1$ \cite{zhong2023extending} & Y & 4.28 & 2.95 & 3.51 & 3.25 & 11.34 &4.28\\ \hline
Single speaker  & Ours & Y & 4.31 & 3.02 & 3.68 & 3.33 & 11.35 & 4.13\\ \hline
Multi speaker & Ours & N &4.22 & 2.75 & 3.36 & 3.19 & 11.35 & 4.14\\ \hline
Multi speaker & Ours & Y & 4.28 & 2.99 & 3.69 & 3.32 & 11.35 & 4.12\\ \hline
Multi speaker+log1p & Ours & N & 4.3 & 2.75 & 3.4 & 3.23 & 11.35 & 4.14\\ \hline
Multi speaker+log1p & Ours & Y & \textbf{4.41} & 3.21 & \textbf{3.78} & \textbf{3.4} & \textbf{11.49} & 4.22\\ \hline

\end{tabular}
\end{adjustbox}
\caption{In-domain Enhancement Results: Valentini \cite{botinhao2016speech}}
\label{tbl:enh-val}
\end{table*}

\begin{table*}
\centering
\begin{adjustbox}{width=1.6\columnwidth}
\begin{tabular}{c c c c c c c c c}
\hline
Pre-train augmentation type & Model & Finetune & CSIG & CBAK & COVL & PESQ & SSNR & NISQA\\ \hline
- & Noisy & - & 2 & 1.52 & 1.6 & 1.47 & 1.42 & 2.79\\ \hline
- & UNIVERSE \cite{serra2022universal} & - & 2.19 & - & 1.76 & 1.51 & - & -\\ \hline
- & SGMSE+ \cite{richter2023speech}  & - & \textbf{2.81} & - & \textbf{2.33} & 1.97 & - & \textbf{4.44}\\ \hline
Random TF mask & \textbf{MAE for restoration}$^1$ \cite{zhong2023extending} & Y & 2.49 & - & 2.05 & 1.81 & - & 2.95\\ \hline
Multi speaker & Ours & N &2.34 & 1.75 & 1.99 & 1.78 & 7.24 & 2.71\\ \hline
Multi speaker & Ours & Y & 2.57 & 2.3 & 2.21 & 2.09 & 8.51 & 3.43\\ \hline
Multi speaker+log1p & Ours & N & 2.45 & 1.8 & 2.1 & 1.8 & 7.24 & 2.74\\ \hline
Multi speaker+log1p & Ours & Y & 2.64 & \textbf{2.51} & 2.32 & \textbf{2.17} & \textbf{9.1} & 3.5\\ \hline

\end{tabular}
\end{adjustbox}
\caption{Out-of-domain Enhancement Results: DAPS \cite{mysore2014can}}
\vspace{-0.16in}
\begin{tablenotes}
   \item $^1$ These numbers in Tables \ref{tbl:enh-val} \& \ref{tbl:enh-daps} were obtained by reproducing \cite{zhang2021restoring} using random TF masking only in pre-training and with STFT.
\end{tablenotes}
\label{tbl:enh-daps}
\vspace{-0.15in}
\end{table*}

\subsection{Finetuning}
\label{sec:finetune}
The pre-trained model can now be finetuned for specific enhancement-related downstream tasks. The fine-tuning model has access to a small amount of paired data. Figure \ref{fig:fine} shows that the fine-tuning model input is the concatenation of the pre-trained model encoder embeddings and the noisy input STFT.
The input data is single-speaker speech corrupted by background noise, background speech and babble, and reverberation and the expected output is clean anechoic speech. The model output is a mask $\in [0,1]$, that computes the amount of clean speech target in each TF bin. This TF mask (not to be confused with pre-training) is multiplied with the noisy input STFT to obtain a clean reconstruction. The noisy phase is used to generate the raw reconstructed waveforms.

The multi-speaker pre-training augmentation also allows for 
source separation downstream task. 
The spectrogram augmentations allow for other downstream tasks like (1) \textit{Bandwidth Expansion}: Bandwidth expansion is facilitated by a pre-trained model that learns to reconstruct masked high-frequencies from frequency masking augmentation. (2) \textit{Packet Loss Concealment (PLC)}: Missing/corrupted speech frames when transported over VoIP can be recovered by time masking augmented pre-training that reconstructs masked time frames \cite{huang2022masked}.
We leave detailed analysis of source separation, bandwidth expansion, and PLC for future work.

\section{RESULTS}
\label{sec:results}
\subsection{Experiment Setup}
\subsubsection{MAE Pre-training}
Pre-training is done on LibriTTS \cite{zen2019libritts} 
using all three train subsets (\textbf{960 hrs}).
Hyperparameters of the ViT-AE pre-training model are from \cite{niizumi2022masked}.
The input audio data is cropped or padded to 4-second chunks at 16 KHz sampling rate. We use an STFT window of 32ms and a hop size of 8ms. ViT-AE is pre-trained for $60$ epochs with a batch size of $256$, using AdamW optimizer with a weight decay of $1e^{-4}$, and learning rate (LR) of $1e^{-4}$ which linearly increases to $1e^{-4}$ during warmup (5 epochs) and then decreases to $1e^{-6}$ by the cosine annealing scheduler \cite{huang2022masked}.
In the augmentation stack, each augmentation is applied with a probability $p$. First, input speech is scaled to a random loudness level $[-30, 10]$ dB, followed by multi-speaker mixture generation (Algorithm \ref{alg:multi}) by picking a secondary interfering speaker from LibriTTS and an RIR from Open SLR28 \cite{ko2017study}.
Next, codec artifacts are added by selecting an audio codec. Random clipping is applied by picking a level $[0,1]$ and finally, background noise is applied by picking a noise from DNS noise \cite{chen2014dns} with a random SNR $[-30,0]$ dB. For spectrogram level augmentation, one of the three masks: time (20\% of time frames are masked), frequency (random high-frequency bins up to 50\% of the bins are masked), and random TF (75\% of TF bins are masked) is chosen for each input clip with probability $p=[0.1,0.1,0.8]$. 


\subsubsection{Finetuning}
The pre-trained ViT encoder is frozen and the encoder patch embeddings are concatenated with the input noisy STFT magnitude after time alignment following \cite{niizumi2022masked}.
Unlike pre-training, the fine-tuning decoder uses global attention, and the batch size is 256,  the LR increases to $2e^{-4}$ during warmup (5 epochs) and decreases to $1e^{-6}$ by 95 epochs of cosine annealing. The model's output is a TF mask that is limited to [0, 1] by a sigmoid function.
We finetune the ViT-AE pre-trained model on the Valentini dataset \cite{botinhao2016speech}  (\textbf{29 hrs}), the test set has 824 noisy-clean paired clips. Zero-shot evaluation is done on the M10 and F10 speakers of DAPS \cite{mysore2014can} dataset which was generated by playing back prerecorded clean speeches in noisy and reverberating environments.
Audio-MAE for restoration \cite{zhong2023extending} is used as the baseline for comparison.

\subsection{Evaluation}
\vspace{-0.5em}
For evaluation we use standard intrusive metrics (PESQ, CSIG, CBAK, COVL, SSNR
\cite{githubGitHubSchmiph2pysepm}) 
and a non-intrusive metric, NISQA \cite{mittag2021nisqa} 
on the Valentini and DAPS datasets. In Table \ref{tbl:enh-val}, we replicate the baseline: \textit{MAE for Audio Restoration} (Table 3 row 4 \cite{zhong2023extending}) using only spectrogram-level 75\% random TF masking for pre-training augmentation. The small discrepancy in numbers is because the baseline uses log-mel spectrograms in pre-training vs STFT used here. We see that when all augmentations (waveform, spectrogram) are applied except multi-speaker mixtures, the fine-tuning performance for in-domain Valentini data is improved. Adding multi-speaker augmentation offers comparable performance to single-speaker augmentation as it is beneficial for removing background babble that is not diffused, background interfering speech, etc. Therefore, full-stack augmented pre-training learns features for reliable application to a broad range of downstream tasks. 
We verify the need for fine-tuning by evaluating directly on the pre-trained model only and observe the dip in performance. Borrowing the idea from Boosting SSL \cite{hung2022boosting}, we replace linear STFT with its $log1p$ compressed features. The input to ViT MAE is $log1p$ compressed version of noisy STFT magnitude and the MSE loss (Figure \ref{fig:pretrain}) is w.r.t $log1p$ features. Table \ref{tbl:enh-val} shows that using $log1p$ compression boosts numbers of both pre-training only and pre-training+finetuning models. Our model outperforms classification-based SSL pre-training \cite{huang2022investigating} and the SOTA method (boosting SSL \cite{hung2022boosting} that uses WavLM \cite{chen2022wavlm}, HuBERT \cite{hsu2021hubert} pre-trained $log1p$ compressed features for finetuning) on 3 out of 4 performance metrics. 
We beat supervised pre-training works like PANNs with UNet \cite{kong2021speech} highlighting the benefits of increased training data availability and diversity (augmentations) from self-supervised learning. Finetuning with multi-speaker augmented pre-training (linear and $log1p$) beats SOTA discriminative models DCUNet \cite{choi2018phase} on all metrics except CBAK. For diffusion-based models like UNIVERSE \cite{serra2022universal} and SGMSE+ \cite{richter2023speech}, our performance is superior for intrusive metrics but underperforms for non-intrusive metrics like NISQA. 


Table \ref{tbl:enh-daps} shows that augmented pre-training offers great performance for out-of-domain generalizations (DAPS \cite{mysore2014can}) over the baseline \cite{zhong2023extending} (Table 5 row 13) and is further improved by using $log1p$ compressed features. Thus, designing an SSL pre-training model that learns distortion removal through augmentation offers robust generalized features for downstream enhancement for both in-domain and out-of-domain data.  
\vspace{-0.25em}
\section{LIMITATIONS \& CONCLUSION}
\vspace{-0.05em}
We designed a universal self-supervised pre-training model that offers robust speech features for various speech enhancement downstream tasks. We show that the distortion removal objective added to MAE pre-training boosts enhancement performance while learning features for other downstream applications. Thus, the pre-trained model can be employed for universal speech enhancement that removes all types of distortions corrupting speech and serves as an enhancement alternative to WavLM. 
We consider enhancement  with background speech removal through multi-speaker augmentations but we hope to explore in the future how the distance-based mixtures lend themselves to the separation of harder mixtures (without distance cues) downstream and zero-shot bandwidth extension and packet loss concealment using spectrogram augmentations of time and frequency masking. 

\vfill\pagebreak